\documentclass[aps,groupedaddress,showpacs,epsfig,floats,12pt]{revtex4}
\usepackage{graphicx}
\begin{document}
\title{Linear magnetic flux amplifier}
\author{D. S. Golubovi\'{c} and V. V. Moshchalkov }
\affiliation{INPAC - Institute for Nanoscale Physics and Chemistry, Nanoscale Superconductivity and Magnetism Group,
Laboratory for Solid State Physics and Magnetism, K. U. Leuven,
Celestijnenlaan 200 D, B-3001 Leuven, Belgium}

\begin{abstract}
By measuring the critical current versus the applied magnetic
field $I_c(\Phi)$ of an Al superconducting loop enclosing a soft
Permalloy magnetic dot, we demonstrate that it is feasible to
design a linear magnetic flux amplifier for applications in
superconducting quantum interference devices. The selected
dimensions of a single-domain Permalloy dot provide that the
preferential orientation of the magnetization is rotated from the
perpendicular direction. By increasing an applied magnetic field,
the magnetization of the dot coherently rotates towards the
out-of-plane direction, thus providing a flux gain and an
enhancement of the sensitivity. As a result of a pronounced shape
anisotropy, the flux gain generated by the dot can be tuned by
adjusting the dimensions of the dot.
\end{abstract} 
\maketitle

\pagebreak

The Superconducting Quantum Interference Device (SQUID) is the
most sensitive magnetic field sensor that has been widely used for
various applications - from the nuclear magnetic resonance
 in medicine to quantum computing \cite{kirt,mooij}.
As the voltage - flux $V(\Phi)$ characteristic  of SQUIDs is
highly nonlinear, they are commonly used in the flux-locked loop,
where a magnetically coupled feedback provides that the SQUID
operates around the steepest part of the $V(\Phi)$ characteristic
\cite{kirt,sci3}. A lot of effort is focused on enhancing the
sensitivity and minimizing the noise in  SQUID systems (see
\cite{sci3} and references therein). Recently, $\pi/2$- and
$\pi$-SQUIDs based on d-wave superconductors have been
investigated as the possible candidates for SQUIDs without the
external flux bias \cite{amin,man,smil}. Soft ferromagnetic
materials have been used as another means to improve the
sensitivity of SQUID magnetometers \cite{xxx,sci2,apl2}. Even
though soft ferromagnets provide an enhancement of sensitivity,
they are generally considered unfavorable because magnetization
switching, domain wall nucleation and motion, as well as
additionally generated vortices can substantially increase the
noise.

Sub-micrometre magnetic elements are in the single-domain state
 and the magnetization reversal occurs through the
coherent rotation \cite{m1,m2}.  By using Landau-Lifshitz-Gilbert
equation the settling time of a sub-micrometre soft magnetic
element in the single domain state is found to be less than
$10\,$ns, which means that such an element does not distort the
applied magnetic field in its vicinity, nor is the magnetization
rotation retarded up to the $GHz$ frequency range \cite{din}.

\begin{figure}[hbt] \centering
\includegraphics*[width=8cm]{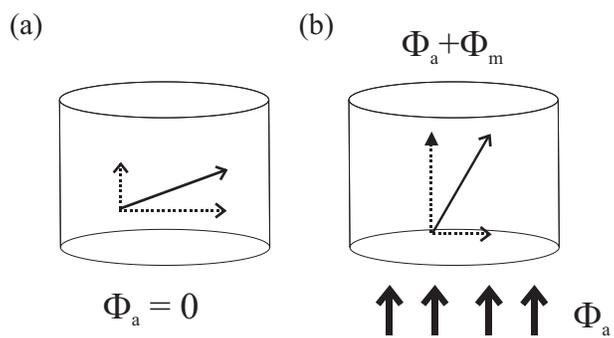}\vspace*{8cm}
\caption{D. S. Golubovi\'{c} and V. V. Moshchalkov} \label{dots}
\end{figure}

We propose to use a soft sub-micrometre magnetic dot as a flux
amplifier for SQUID applications. If the height of a soft dot is
slightly smaller than its diameter, the magnetization is oriented
neither in-plane nor out-of-plane, as shown in Fig. \ref{dots}(a)
\cite{m2}. By applying a perpendicular magnetic field, the
out-of-plane component of the magnetization increases and,
accordingly, the total flux generated by the dot along the
out-of-plane direction increases, as shown in Fig. \ref{dots}(b).
If such a dot is placed at the centre of the pick-up loop of a
SQUID, it can act as a built-in flux amplifier.

\begin{figure}[hbt] \centering
\includegraphics*[width=6cm]{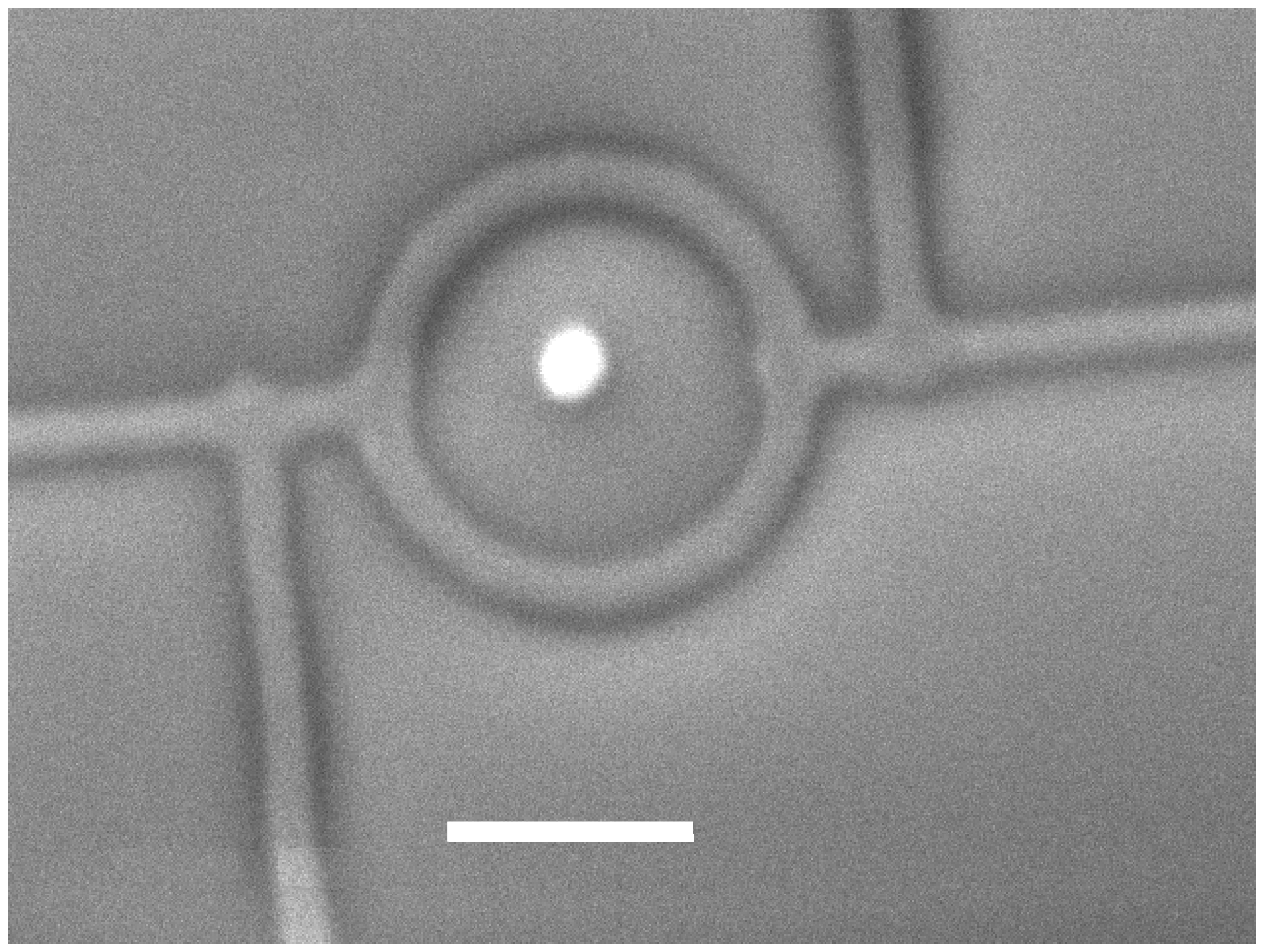}\vspace*{8cm}
\caption{D. S. Golubovi\'{c} and V. V. Moshchalkov} \label{sam}
\end{figure}

\begin{figure}[hbt] \centering
\includegraphics*[width=8cm]{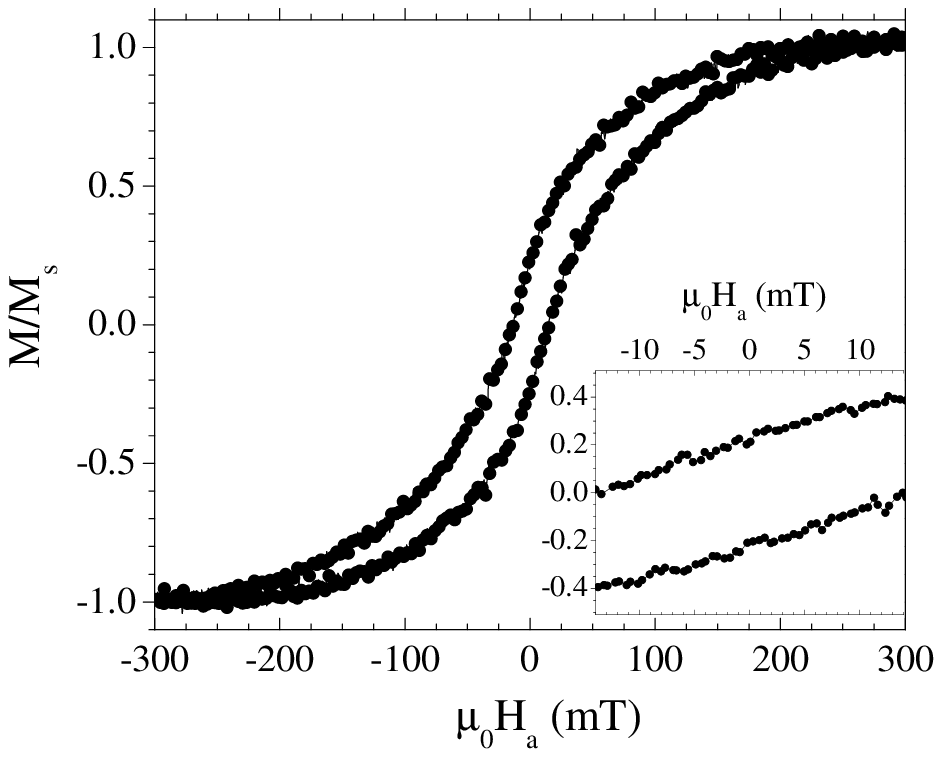}\vspace*{8cm}
\caption{D. S. Golubovi\'{c} and V. V. Moshchalkov} \label{hyst}
\end{figure}

The properties of the flux amplifier have been studied by
measuring the critical current versus the applied magnetic field
$I_c(\Phi)$ of a superconducting loop with a soft magnetic dot at
its centre. The samples were prepared using electron beam
lithography and lift-off procedure in two steps. In the first step
$45\,$nm thick Al loops were prepared by thermal evaporation,
whereas in the second step $90\,$nm thick polycrystalline
Ni$_{0.8}$Fe$_{0.2}$ Permalloy (Py) dots were grown by electron
beam evaporation. Two samples hereafter referred to as the 'Sample
A' and 'Sample B', as well as a reference Al loop without a Py dot
have been measured. Fig. \ref{sam} shows a scanning electron
micrograph of the Sample B. The white
 bar corresponds to $1\,{\rm \mu m}$. The hysteresis loop of an array
 of the co-evaporated reference  Py dots measured at $5\,$K in the perpendicular magnetic field is shown in
 Fig. \ref{hyst}. The inset in Fig. \ref{hyst} shows that the magnetization of the
dots depends linearly on the applied field in the range
$|\mu_0H_{a}|\leq 15\,$mT.  The inner and the outer radii of
 the Sample A are $r_{iA}=0.73\,{\rm \mu m}$ and $r_{oA}=0.94\,{\rm \mu
 m}$, the radius of the Py dot is $r_{dA}=0.12\,{\rm \mu m}$, whereas for
 the sample B $r_{iB}=0.73\,{\rm \mu m}$, $r_{oB}=0.92\,{\rm \mu
 m}$ and $r_{dB}=0.115\,{\rm \mu m}$. The radii of the reference sample are $r_{iR}=0.73\,{\rm \mu m}$
 and $r_{oR}=0.95\,{\rm \mu m}$, respectively. The critical temperatures of the Samples A
 and B are $T_{c0A}=1.4571\,$K and $T_{c0B}=1.4557\,$K, whereas the critical temperature of the
 reference sample is $T_{c0R}=1.4591\,$K. The
 superconducting coherence length is $\xi(0)=117\,$nm and the
 penetration depth is $\lambda (0)=374\,$nm or $\kappa \approx 3.2$.
 Since at the temperatures close to the zero-field
 critical temperature $T_{c0}$ the superconducting coherence length
 $\xi(T)$ is greater than the radius of the loop, the presence of the
 non-current-carrying voltage contacts creates two parallel weak
 links and the $I_c(\Phi)$ of the loop has the same periodicity as
 the $I_c(\Phi)$ of the SQUID \cite{vvm}.

The measurements have been carried out in the DC mode, with the
current and field steps of $10\,$nA and $50\,{\rm \mu T}$,
respectively. The field and transport current were set in the
order which ensures that no flux can be trapped in the samples
over the course of the measurements.

\begin{figure}[hbt] \centering
\includegraphics*[width=8cm]{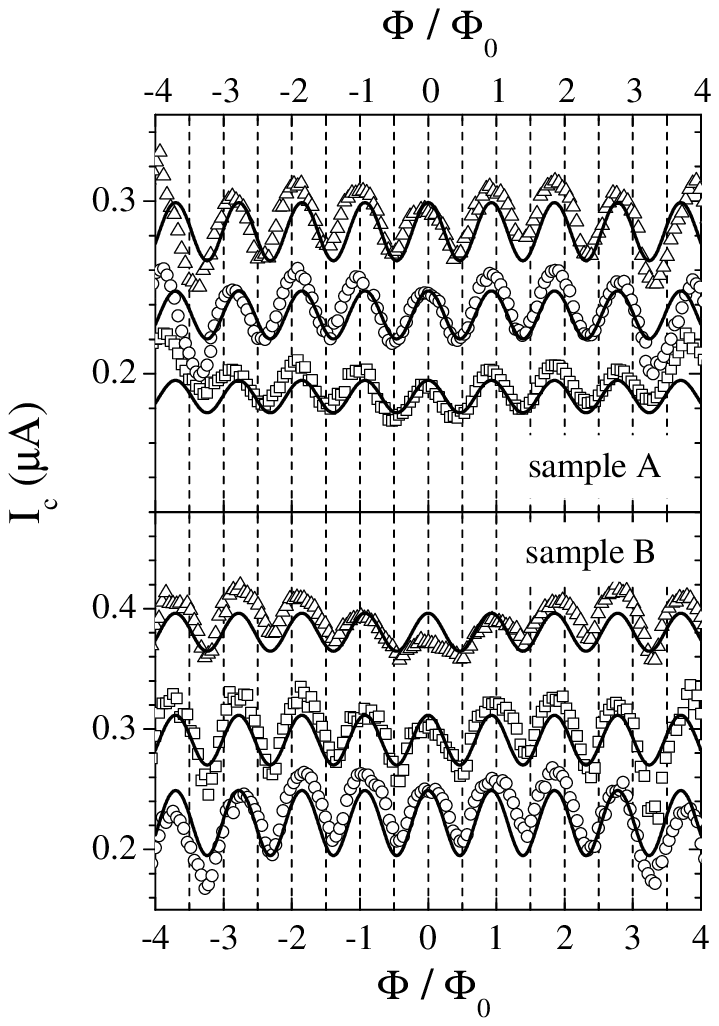}\vspace*{8cm}
\caption{D. S. Golubovi\'{c} and V. V. Moshchalkov} \label{icba}
\end{figure}

The upper part of Fig. \ref{icba} shows the $I_c(\Phi)$-curves of
the Sample A taken at $0.997T_{c0A}$, $0.995T_{c0A}$ and
$0.993T_{c0A}$, whereas the lower part shows the
 $I_c(\Phi)$-curves of the Sample B at $0.995T_{c0B}$,
$0.994T_{c0B}$ and $0.992T_{c0B}$. A higher critical current
corresponds to a lower temperature.   The critical currents were
determined using the conventional $1\,{\rm \mu V}$-criterion.

\begin{figure}[hbt] \centering
\includegraphics*[width=8cm]{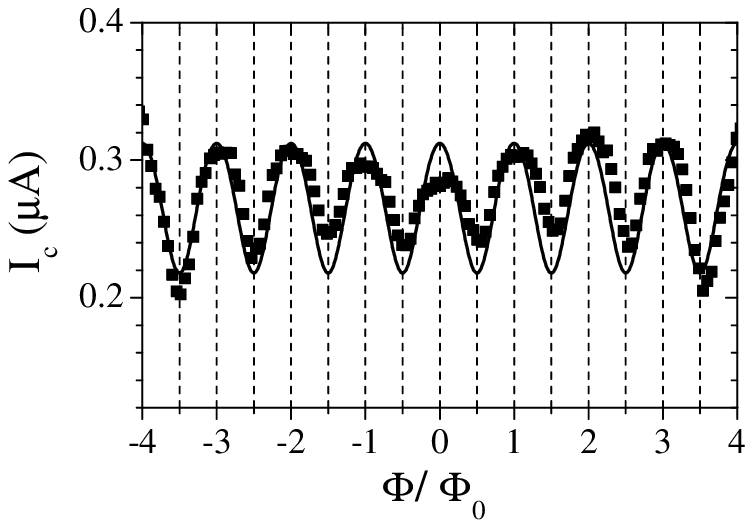}\vspace*{8cm}
\caption{D. S. Golubovi\'{c} and V. V. Moshchalkov} \label{icbref}
\end{figure}

Open symbols in Fig. \ref{icba} are the experimental data, whereas
the solid lines are the theoretical curves, obtained using the
expression for the critical current of an asymmetric SQUID,
modified to take into account the influence of the soft dots
\cite{smil}
\begin{equation}
I_c(\Phi)=\left
(A+B\cos\left[2\pi\left({\Phi\over{\Phi_0}}+g_{m}{\Phi\over{\Phi_0}}\right)\right]\right)^{1/2}
\label{ic}
\end{equation}
whereby $A$ and $B$ are constants with the dimension [$A^2$],
$\Phi $ is the applied flux, $\Phi_0$ is the superconducting flux
quantum and $g_m$ is the gain provided by the magnetic dot. Given
the linear dependence of the magnetization on the applied magnetic
field (see Fig. \ref{hyst}), the flux gain provided by the dots
has been taken constant. The flux $\Phi $ has been calculated with
respect to the means radius $r_m=(r_o+r_i)/2$, as at the lowest
temperature measured of $0.992T_{c0B}$ the following condition is
valid $\xi(T)>>r_o-r_i$ (($\xi(0.992T_{c0B})\approx 1.3\,{\rm \mu
m}\, , \,r_{oB}-r_{iB}=0.187\,{\rm \mu m}$). The 1D-character of
the order parameter in the loops is of an extreme importance for
the validity of the measurements, because it ensures that the
radius
 over which the fluxoid quantization occurs
does not change with the temperature and/or field, thus ruling out
any inherent change in the periodicity of the $I_c(\Phi)$-curves
\cite{den}. The constants $A$ and $B$ have been chosen so as to
provide the best amplitude agreement, whereas for both samples,
irrespective of the temperature, the gain provided by the dot is
$g_m=0.1$, which means that for this particular dimensions the Py
dot provides a $10\,\%$ enhancement of the sensitivity. For
comparison, Fig. \ref{icbref} shows the $I_c(\Phi)$-curve of the
reference sample taken at $0.994T_{c0R}$. Filled symbols are the
experimental data, whereas the solid line is the theoretical curve
obtained by setting $g_m=0$ in Eq. (\ref{ic}). It is clear that
without a Py dot, the $I_c(\Phi)$-curve has the periodicity
$\Phi_0$, independent of an applied magnetic field.

A moderate gain of only $10\,\%$ is caused by the low thickness
(only $150\,$nm) of the electron beam resist, which limited the
maximum achievable height of Py dots and, in turn, constrained the
radius of the dot. An increase and adjustment of the gain can be
accomplished by using thicker resists, which would allow to
increase the height and radius of a Py dot, whilst keeping the
height/diameter ratio which provides a rotated preferential
direction of the magnetization.

\begin{figure}[hbt] \centering
\includegraphics*[width=6.5cm]{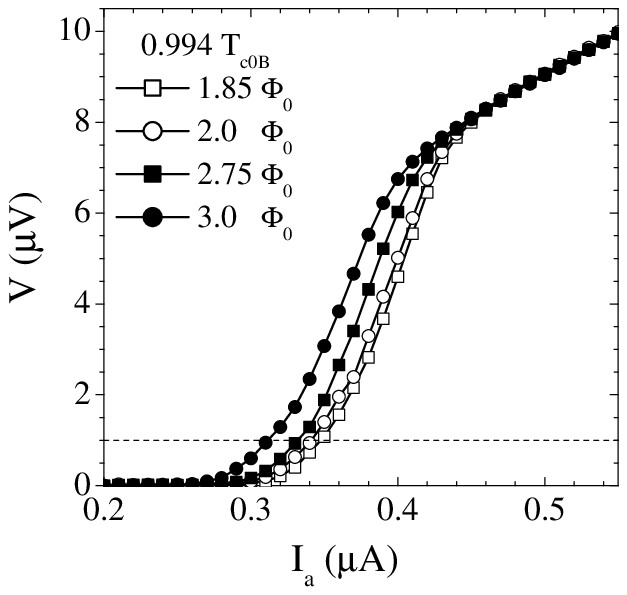}\vspace*{8cm}
\caption{D. S. Golubovi\'{c} and V. V. Moshchalkov} \label{iv}
\end{figure}

Fig. \ref{iv} shows the voltage-current (IV) curves of the Sample
B at $1.85\Phi _0$ and $2\Phi _0$ (open symbols), and $2.75\Phi
_0$ and $3\Phi _0$ (filled symbols) taken at $0.994T_{c0B}$. The
horizontal lines indicate the voltage criterion used. Non-integer
values of the flux in Fig. \ref{iv} correspond to the maxima in
$I_{c}(\Phi)$-curves shown in Fig. \ref{icba}. The IV-curves
explicitly show that as the applied magnetic field increases, the
difference between the flux values whereat a maximum in the
critical current appears and the closest integer flux ($n\Phi
_{0}\, , \, n\in {\bf Z}$) increases. We note that neither local
heating nor non-equilibrium effects have been observed in the
IV-curves \cite{gol}. This clearly implies that the contraction of
the
 period of the $I_{c}(\Phi)$-curves comes from the magnetic dot.

In conclusion, we have fabricated and investigated a magnetic flux
amplifier for SQUID applications. By measuring $I_c(\Phi)$-curves
it has been demonstrated that a Py magnetic dot provides a linear
flux amplification. The design proposed makes it possible to tune
the flux gain by adjusting the height/diameter ratio of a Py dot.

This work has been supported by the Research Fund K. U. Leuven
GOA/2004/02 programme, the Flemish FWO and the Belgian IUAP
programmes, as well as by the JSPS/ESF "Nanoscience and
Engineering in Superconductivity" programme.

\pagebreak
\begin{center}
\Large{Figure Captions}
\end{center}
\vspace*{2cm}
\begin{description}
\item[Fig. 1]{A schematic of the interaction of a soft magnetic
dot with an applied magnetic flux.} \item[Fig. 2]{A scanning
electron micrograph of an Al loop with a Py dot at the centre
(Sample B). The white bar corresponds to $1\,{\rm \mu m}$.}
\item[Fig. 3]{The hysteresis loop of Py dots used in the
experiment measured at $5\,$K in the perpendicular field. The
inset shows the magnetization of the dots for the applied fields
$|\mu_0H_a|\leq 15\,$mT.} \item[Fig. 4]{The $I_c(\Phi)$-curves of
the Sample A (upper part) taken at $0.997T_{c0A}$, $0.995T_{c0A}$
and $0.993T_{c0A}$ and  Sample B (lower part) taken at
$0.995T_{c0B}$, $0.994T_{c0B}$ and $0.992T_{c0B}$. Open symbols
are the experimental data, whereas the solid lines are the
theoretical curves. A higher critical current corresponds to a
lower temperature.} \item[Fig. 5]{The $I_c(\Phi)$-curve of the
reference sample taken at $0.994T_{c0R}$. The filled symbols are
the experimental data, whereas the solid line is the theoretical
curve.}\item[Fig. 6]{The $IV$-curves of  the Sample B at
$0.994T_{c0B}$ for $\Phi =1.85\Phi_{0}$, $\Phi =2\Phi_{0}$, $\Phi
=2.75\Phi_{0}$ and $\Phi =3\Phi_{0}$.}
\end{description}

\pagebreak \vspace*{3cm}
\pagebreak \vspace*{3cm}
\pagebreak \vspace*{3cm}
\pagebreak \vspace*{3cm}
\pagebreak \vspace*{3cm}
\pagebreak \vspace*{3cm}

\end{document}